# Monte-Carlo Modelling of the Electron Spectra of $^{235}$U- and $^{239}$Pu- Films, Irradiated by Thermal Neutrons, Due to All Possible Mechanisms Excluding β-Decay.
# Comparison With Experiment

V. D. Rusov$^a$, V. N. Pavlovych$^b$, V. A. Tarasov$^a$, S. V. Iaroshenko$^b$, D. A. Litvinov$^a$

$^a$Odessa National Polytechnic University, Odessa, Ukraine
$^b$Institute for Nuclear Research NAS of Ukraine, Kiev, Ukraine

The electron energy spectra, not connected to β-decay, of $^{235}$U- and $^{239}$Pu-films, irradiated by thermal neutrons, obtained by a Monte-Carlo method is presented in the given work. The modelling was performed with the help of a computer code MCNP4C (Monte-Carlo Neutron Photon transport code system), allowing to carry out the computer experiments on joint transport of neutrons, photons and electrons. The experiment geometry and the parameters of an irradiation were the same, as in [11] (the thickness of a foil varied only). As a result of computer experiments, the electron spectra was obtained for the samples of $^{235}$U, $^{239}$Pu and uranium dioxide of 93 % enrichment representing a set of films of 22 mm in diameter and different thickness: 0,001 mm, 0,005 mm, 0,02 mm, 0,01 mm, 0,1 mm, 1,0 mm; and also for the uranium dioxide film of 93 % enrichment (diameter 22 mm and thickness 0,01mm), located between two protective 0,025 mm aluminium disks (the conditions of experiment in [11]) and the electron spectrum was fixed at the output surface of a protective disk. The comparative analysis of the experimental [11] and calculated β--spectra is carried out.

## 1. Introduction

The interpretation of antineutrino experiments at nuclear reactors is impossible without knowledge of a total antineutrino energy spectrum ("at the birthplace") of the fission product mixture, which are formed in the reactor core, and moreover without a priory knowledge of antineutrino energy spectrum of each fissile isotopes - $^{235}$U, $^{238}$U, $^{239}$Pu, $^{241}$Pu - components of nuclear fuel [1-3].

The most comprehensive and physically proved characteristic of the calculated ν-spectra of the β-active nucleus is it total emission β-spectra obtaining (experimentally or by calculation) by weighing of partial β-spectra of permitted and forbidden β-transitions [4]. The procedure of obtaining of the total ν-spectra from the total β-spectra of the given nucleus is based on the kinematical connection between electron and antineutrino: $E_ν = E_0 - E_β$, where $E_0$ is the total energy, which is distributed between electron ($E_β$) and antineutrino ($E_ν$). In other words, here the law of energy conservation determines a way of a calculation of any partial transition ν-spectrum: the last is symmetric to a partial β-spectrum, i.e. $N_ν(E_ν) = N_β(E_β - E_ν)$. This circumstance not only allows to obtain a ν-spectrum but, that is most important to obtain it with accuracy equal to that of calculated and/or measured β-spectrum. Therefore, the reliability of the forecast of an effective total ν-spectrum of each fissile isotope of nuclear fuel is defined by reliability of the used catalogue of the total β-spectra of the given fission products, and by the correctness of the description of these fission product accumulation [4].

Alongside with high quality of ν-spectra theoretical calculation, it is necessary to note also the strongly increased level and certain successes of experimental ideology and technique of neutrino measurements. For example, now there is a well-developed method (introducing by the group of Mikaelian from Kurchatov institute [5]) of direct measurements of electronic antineutrino energy spectra, emitted by an a priory known composition of the basic fissile isotopes in reactor experiments. On the basis of the experimentally measured β-spectra of $^{235}$U [6], $^{239}$Pu [7] and $^{241}$Pu [7] fission products, the converted ν-spectra of the specified isotopes were obtained. In the case of $^{238}$U, for which the β-spectrum was not measured, the calculated spectrum of electronic antineutrino is usually used [8].

It is natural, that the requirement to accuracy of ν-spectra determination (and therefore to adjusted β-spectra) is increased with time. That is due to transition from qualitative experiments to finding - out of quantitative laws, extremely important both for research of neutrino physics fundamental problems, for example, neutrino oscillation, estimation of neutrino weight, estimation of a constant of weak interaction etc. [9,10], and for the practical task solution, for example, effective solution of the problem of intrareactor process neutrino diagnostics [1,2].

The analysis of the reasons of distortion of experimentally measured spectra of β-particles taking off from the targets of uranium and plutonium, specifies the significant fact, that, except for the classical corrections to a spectrum connected with scattering and electron energy losses in a target material, it is necessary also to take into account the essential influence of an additional electron source, not connected to β-decay of fission products but caused by the interaction of target atoms with γ-radiation created during neutron transport. Some influence may be caused also by the secondary electrons due to electron-electron scattering.

The evaluation shows that the problem of difference between the β-spectra "at the birthplace" and β-spectra "at the place of an output" from the fissile material can not be solved by the experimental methods [11]. But today the hope is born that due to fast developing of the computers and special computer codes the β-spectra "at the birthplace" one can obtain by the modelling codes, for example, similar to [12]. Some of the today codes allow the real possibility to perform the computer experiment to measure not connected to β-decay electrons from the $^{235}$U - and $^{239}$Pu – films irradiated by the thermal neutrons. That is the main aim of the given work.

## 2. The method of experimental measurement of β-spectra and its conversion to electronic antineutrino energy spectra.

The method of determination of the electronic antineutrino energy spectra of the fission products of $^{235}$U on the basis of an experimental β-spectrum for the first time was applied in [13]. The method assumes the obtaining of a ν-spectrum from an experimental β-spectrum by the integral equation solving. As the analysis of work [4] shows, the method of [13] brings the significant uncertainty to a restored ν-spectrum of a of the fission product mixture. This uncertainty is caused by the assumption that all β-transitions in a fission product mixture are solved, and also by the procedure of averaging on the charge values and computing difficulties of threefold differentiation. However, considering a modern level of various libraries of the nuclear data (for example, [14] and [15]), Russian catalogues of the total β, ν-spectra of the radioactive isotopes in a wide range of mass numbers [4], the effective development of solving methods of the incorrect problems [16], to which the basic integral equation of a method relates [13], and the quality of modern mathematical packages, it is possible to hope, that these arguments now can practically be removed.

Within the framework of a discussed problem of determination of a ν-spectrum from an experimental spectrum of β-particles, emitted by fission product mixture, a so-called method of the steady ratios [8,17] and method of effective summation [4,18] are most frequently applied at the present time. A method of the steady ratios has arises as a result of numerical modelling of the $\beta$, ν-spectra of the fission fragments, which has allowed to find the simple algorithm connecting the antineutrino spectrum $\rho_v$ and beta-spectrum $\rho_\beta$, which are formed simultaneously during the decay of fragments. Besides, the procedure of $\rho_v$ obtaining «at the birthplace» for the given fissile nucleus is based on the fact, that the ratio k (E) = $\rho_v$ (E) /$\rho_\beta$ (E) in equilibrium condition has a good stability, which is not depended on the hypotheses of unknown decay scheme. Although the stability of this ratio has no clear physical base [4], the experimental data of measurements of fission fragment β-spectra of $^{235}$U, $^{239}$Pu and $^{241}$Pu have confirmed the results of modelling [8,17] with a good accuracy. In a case of $^{238}$U, for which the measurement of a β-spectrum is absent, one can use as $\rho_v$ the spectrum [8] founded by the calculations and corrected with the ratio k (E) for $^{235}$U.

As an algorithm of transition from the β-spectrum to the antineutrino spectrum, the method of direct summation of β, ν-spectra of the given nuclei forming the fission product mixture [4] is used taking into account the concrete conditions of fuel irradiation in nuclear reactor, and corresponding effective total β, ν-spectrum is:

$$N_{eff}^{\beta,v}(E,U,T,t) = \sum_j Q_j(U,T,t) \cdot N_j^{\beta,v}(E) \tag{1}$$

Here $Q_j(U, T, t)$ is the activity of the j-th nucleus depending on the conditions of fuel irradiation (constituents of initial fuel, density of a neutron flux, burn-up and other parameters influent on the every j-th nucleus accumulation); T, t are the time of fuel irradiation and time of endurance after a previous irradiation; $N_j^{\beta,v}$ (E) - total β- or ν-spectrum of j-th nucleus. In most cases the calculated or experimental effective total spectrum $N_{eff}^{\beta,v}$ (E, U, T, t) is measured in units (keV·fis)$^{-1}$, that, in turn, corresponds to concentration $Q_j$ (U, T, t) calculation per fission rate.

The *apriori* knowledge of $N_i(t)$ and $\Phi$ values (the concentration of the i-th fissile isotope and neutron flux, correspondingly) determines the opportunity of j-th fission product presence in local volume of the reactor core by the direct solution of the differential equation describing the change in time of the j-the nucleus concentration $C_j(t)$ [4]:

$$\frac{dC_j(t)}{dt} = -\lambda_j C_j(t) + \sum_{i=1}^{p} \langle y \rangle_{ij} \langle \sigma_f \rangle_i \Phi N_i(t) + \sum_{m=1}^{j-1} \lambda_{mj} C_m(t) + \sum_{m=1}^{j-1} \langle \sigma_c \rangle_{mj} \Phi C_m(t) \quad (2)$$

with initial condition $C_j(0) = C_{0j}$.

Here the index m belongs to the nucleus - predecessor, $m < j$; $<y>_{ij}$ is the independent yield of the j-the nucleus due to fission of the i-th component of fuel averaged by the neutron spectrum; $\lambda_{mj}$ is the decay probability of m-th nucleus to j-th one by any way; $<\sigma_f>_i$ is the one-group fission cross section of the i-th fissile nucleus; $<\sigma_c>_{mj}$ - one-group cross section of reactions (n, γ), (n, 2n) of the m-th nucleus with formation of j-th one. Let's note, that the index i varies within the limits $1 < i < p$, and index j runs over all known fission products, for example, Russian catalogue gives $1 < j < 1028$ [4] So one should solve the huge system of the differential equations. The equation (2) is wrote in one group approximation, but it is possible, of course, to solve many group equations. The concentrations $C_j(t)$ are connected with j-th nucleus activity as follows: $Q_j(t) = \lambda_j \cdot C_j(t)$. The way of system

(2) solution is described, for example in [19] and is realised in many known computer codes.

However it is very important to note, that all three considered above methods of obtaining of the ν-spectrum from an experimental β-spectrum, namely, method of Reines et al. [13] and method of the steady ratios [8,17], generally in a combination with a method of direct summation [4], comprise the experimentally ineradicable contradiction. Although the methods are intended for the antineutrino energy spectra determination «at the birthplace, its use the experimental β-spectrum «at the place of registration», that can essentially deform a required antineutrino spectrum due to the strong change of electronic spectra «at the birthplace» in comparison with electronic spectra «at the place of an output» from the fissile material during the electron and neutron transport in condensed medium. However, because of the absence of experimentally checked alternatives, it is necessary to reconcile with this contradiction, but it is not necessary to forget about it during the error estimation of the spectra, obtained by this method.

## 3. Experimental measurement of β-spectra emitted by the fission fragments

It is obvious, that the aspiration to decrease the difference between the experimental **β**-spectrum «at the place of registration» and **β**-spectrum «at the birthplace», should put the appropriate rigid requirements to a technique of experimental measurements. To avoid the strong change of electronic spectra «at the birthplace» in comparison with electronic spectra «at the place of an output» from fissile material, the thickness ($\ell$) of the electron source (usually the foil of appropriate fissile isotope, irradiated with thermal neutrons) should be smaller than the free path length of the electros of minimal energy, that is, $\ell < \lambda_\beta^{E_{min}}$. On the other hand, the thickness of a foil should be greater than the free path length of thermal neutrons for fission reaction ($l_f = 1/\Sigma_f$, where $\Sigma_f$ is the fission macroscopic cross section), that is, $\ell > l_f$. The number of fissions in a source otherwise decreases, that reduces the electron flux. It is impossible to achieve the full coincidence of a $\beta$-spectrum «at the birthplace» with a $\beta$-spectrum «at the place of an output» and moreover «at the place of registration» by foil thickness decreasing due to electron electromagnetic interaction with a medium. However, it is possible to use the following estimations of average path lengths of monoenergetic electrons ($R_\beta^{E_\beta}$) in $^{235}$U and $^{239}$Pu for the practical choice of foil thickness. According to [20], for $^{235}$U $R_\beta^{E_\beta=0.1 MeV} \approx 0,01$ mm, $R_\beta^{E_\beta=1.0 MeV} \approx 0,4$ mm, $R_\beta^{E_\beta=3.0 MeV} \approx 1,0$ mm. One can obtain practically the same estimations for $^{239}$Pu. Taking into account the estimation $l_f \approx 0,35$ mm for $^{235}$U, we can conclude that to measure the electron spectra with energy greater than 1.0 MeV, the compromise choice of foil thickness is $\ell \sim 0,35$ mm. To measure the electron spectra of significantly smaller energy than 1.0 MeV, the compromise choice of foil thickness is $\ell < 0,01$ mm, since it is impossible to satisfy both requirements. It should result in significant decreasing of fission number in a source, that reduces an electron flux. Probably, it is possible to compensate it by the increase of thermal neutron flux, or significant increase of time of measurements.

In this sense, it is necessary to mark the work [21], in which the reasons of distortions of experimental spectra of *β*-particles were analysed. It is marked, that the basic reasons of distortions of a spectrum of *β*-particles which are emitted by the targets of uranium and plutonium are: a) scattering and electron energy losses in a material of a foil and cladding; b) the same in the air and films during the particle flight to spectrometer and c) the distortions in spectrometer. The corrections of spectra caused by the reasons b) and c), are small. The measurements with thin sources have shown, that the difference between the experimental and theoretical form of a spectrum does not exceed 4-5 % for the energies E > 1 MeV. Some methods and the procedure of these correction determinations, consisting in accounting for the spectrometer response function for thin sources, were studied in works [11,22-23], which details will be considered below. The appreciably large corrections were caused by the electron scattering and energy losses in targets (the reason a)). Those were determined in the measurements with various electron sources ($^{207}Bi$, $^{56}Mn$, $^{144}Ce$, $^{144}Pr$, $^{42}K$, $^{38}Cl$, $^{252}Cf$), each of which was located between two lead films and into the package for targets. Thus, the electron output from a uranium and plutonium material was simulated, and the appropriate corrections were determined. The significant at small energies corrections were quickly decreased with energy increase (10 % at E=3 MeV). These corrections, according to measurements and calculations, have weak dependence on the nature of the electron source. The results of correction measurements caused by the target thickness, material of packing and so on (reason a)), are presented at Fig. 1.

In general, it is necessary to note, that there are a little works in this area, apparently, because of complexity, labour consume and dearness of experiment realisation. The first work in this area was [13], in which the electron spectrum was measured in the energy range 1.5 MeV - 8 MeV (high energy part of a spectrum) emitted by $^{235}$U foil (19.05 mm in diameter and with thickness 0.003 mm), which was located between two aluminium disks of the same diameter and with thickness 0.05 mm, irradiated with a beam of reactor thermal neutrons with flux $10^6$ n / (cm$^2$·s). Then the works [11, 24] were published, which in many respects of technique repeated the work [13]. In [24], the foil of $^{235}$U was also a source of electrons (diameter 22 mm and thickness 0,005 mm, thickness of a protective aluminium disk 1,0 mm), irradiated by reactor thermal neutrons (with a flux $10^7$ н / (cm$^2$s)). The electron spectrum was measured in energy range 0,75 MeV - 8 MeV, the error of measurements was estimated as ~5 % at steady reactor operation and ~8 % at pulse regime. In [11], the UO$_2$ foil of 93 % enrichment (diameter 22 mm and thickness 0,01, thickness of a protective aluminium disk 0,025 mm), irradiated by thermal neutrons (flux 3·105 н/(см$^2$·с)) was a source of electrons. In this work, the total electron spectrum in the energy range 0,1 MeV - 8 MeV was measured, the error of measurements was estimated as less than ~ 7 %. The results of measurements are presented at fig. 2. It was marked in work, that the total electron spectrum was measured: the β-particles and conversion electrons, and it was assumed, that 95 % of conversion electrons have energies < 0,2 MeV, and their contribution can be neglected.

Let's note, that in the works [11,24] the time dependence of the electron spectrum approach to an equilibrium spectrum, that is, the dependence of a spectrum on time, past from the moment of a neutron irradiation beginning, and number of other temporary parameters, in particular, the change of a spectrum at a pulse irradiation of a sample, and the change of a spectrum after an irradiation also was investigated.

The measurement of spectra of the electrons emitted by the fissile isotope irradiated with thermal neutrons is connected to the certain experimental difficulties. During the transition from energy 1 MeV to 8 MeV the number of β-particles decreases on three order of magnitude, and the relatively small discrepancy in energy scale graduation of the detector results in appreciable distortion of a spectrum.

The difficulties arise also in connection with necessity of background suppression, since the background component always exists in such experiments due to fission processes (prompt and delayed neutrons and γ-quanta), to scattered neutrons and γ-quanta and induced radioactivity of experimental facility constructional elements etc.

In the reviewed works the gradient technique of a background suppression designated as (ΔE/Δx)E was used, where the system of electron registration consists of two detectors: flying and basic. The flying detector containing small quantity of substance (for example, in [13] the gas proportional counter was used) is usually located in front of the basic detector and registers flying it electrons with practically 100 % of probability. Besides, the flying detector was poorly sensitive to γ-radiation and, apparently, to neutral particles. The basic detector is intended for electron spectra measurements. The detectors were included in coincidence scheme, that allow to suppress a background in tens time. For measurement and subsequent account of the background created by an induced radioactivity of constructional elements, in some works, for example in [11], the following was made. In experimental facility consisting usually of two chambers (in one of which the electron source is placed, and in another the detectors) the electrons get to the chamber with detectors through a narrow entrance aperture, which in case of necessity can be blocked by a screen. Therefore, the background created by the induced radioactivity of constructional elements can be measured at blocked by a screen electron entrance.

The standard sources of conversion electrons were used to graduate the system: $^{137}$Cs (624 keV), $^{207}$Bi (482 keV and 991 keV) and some others. Since the energy spectra are usually represented in units 1/(MeV·fis), the monitoring of fission number in a sample was carried out by the special detector placed on a surface of a sample, opposite to that of detector location.

In [25] with the help of γ-spectrometer the force function of β-decay of about 50 neutron excess fission products separated by mass separator was measured. In work [26] the spectrum of reactor antineutrino calculated by computer processing of experimental data, which were obtained in [25] for the given neutron surplus nucleus and their subsequent summation.

On 80th the works [21,27,28,6] were published. In work [21] the measurement of the ratio of spectra $\rho_\beta^5(E)/\rho_\beta^9(E)$ and its comparison with calculated values [29] was carried out.

For this purpose the facility "Disk" was specially designed. On a rim of a rotating organic glass disk the rectangular slices 20x30 mm of $^{235}$U and $^{239}$Pu metal foil by thickness 0,02 mm settled down. Third of disk rim was occupied by uranium target, second third – by plutonium film, the last third was used for background measurements. The disk rotated with rate 13 s$^{-1}$. The source of neutrons, which was the ampoule containing $^{252}$Cf with full neutron flux ~10$^7$ n/s, was settled down from the one side of an axis, the β-spectrometer was located from other side of an axis. The lead and boron polyethylene protection was located between a source and spectrometer. The background, as well as in the previous experiments, was eliminated using the method (ΔE/Δx)E, that is, use of two detectors: flying detector made of a thin plate of scintillating plastic, and basic (stilbene crystal by a diameter 5 cm and thickness 5 cm), switched in the coincidence circuit. At this facility the measurement of spectra $\rho_\beta^5(E)$ and $\rho_\beta^9(E)$ was carried out practically simultaneously. The experimental values of $\rho_\beta^5(E)/\rho_\beta^9(E)$ (fig. 3) were obtained which have coincided with calculated values within the limits of an experimental error of.

Essential distinction (especially significant in high energy part) of β-spectra $\rho_\beta^5(E)$ and $\rho_\beta^9(E)$ for the first time was experimentally confirmed, and hence the corresponding antineutrino spectra, that experimentally confirm the necessity of the account of change of the isotope content of the core during the neutrino reactor experiments.

In works [27,28,6] the results of measurements of electronic spectra due to fission of $^{235}$U, $^{239}$Pu, $^{241}$Pu and corresponding antineutrino energy spectra, obtained at Grenoble research reactor, are presented. There the source of electrons (target for neutrons) consists of a fissile material foil of 3x6 cm with thickness 0,0005 mm and protective Ni foils of thickness 7 mg/cm$^2$. The measurements were carried out by magnetic β-spectrometer, removed from a source on distance ~ 13 m. The technique (ΔE/Δx)E using multifilament proportional counters and plastic stintillator was applied also. The background was determined in experiments, in which the target fissile material was replaced by nonfissile one of equivalent sizes and weight. In order to define the number of β-particles $\rho_\beta(E)$ emitted in one fission act in an energy interval E, E + ΔE, instead of a target of fissile substance the control foil of substances with well known neutron capture cross sections and known factors of conversion α were located. In this case

$$\rho_\beta(E) = \frac{N_f(E)}{N_{conv}(E)} \cdot \alpha \cdot \frac{k\sigma_{n\gamma}n}{\sigma_{nf}n_f}, \qquad (3)$$

Where $N_f(E)$ is the number of electrons registered in the energy interval ΔE during the measurements with a fissile foil, $N_{conv}(E)$ is the same value for conversion electrons, $k$ is the yield of γ-quanta of the given energy at neutron capture, $\sigma_{nf}$ is the fission cross section, $\sigma_{n\gamma}$ is the capture cross section, $n_f$ is the number of fissile atoms, n is the same value for a control foil.

The errors of results were estimated ∼ ±3 %. The results of measurements are presented at fig.4.

## 4. Electron spectra not connected to β-decay from $^{235}$U- and $^{239}$Pu-films irradiated by thermal neutrons.

In the given work the model Monte Carlo electron energy spectra, not connected to β-decay, from $^{235}$U- and $^{239}$Pu films, irradiated by thermal neutrons, and their comparison with experimental data [11] are presented The modelling was made with the help of a computer code MCNP4C (Monte-Carlo Neutron Photon transport code system), allowing to carry out the computing experiments on modelling of the joint transport of neutrons, photons and electrons. MCNP4C allows to obtain the estimations of the energy spectra of the electrons, coming out of the sample surface, in which one modes the transport of neutrons. The mechanisms of electron formation, realised in the code, are following: nucleus fission and the radiation capture of neutrons are accompanied by the γ-quantum emission, which during the transport in a sample create the electrons on the basis of a classical triad of processes of γ-quantum interaction with a matter (formation of the electron-positron pairs, emission of the photoelectrons, Compton effect and emission of the Auger electrons). The movement and energy of every electron is traced up to the moment of their exit through a surface of a sample. The sizes, element contents of a sample and the characteristics of a neutron source corresponding to the irradiation conditions of experiment, are set in the input file of given task. As such electrons are not connected to the processes of β-decay but give the contribution to resulting β-spectra measured in experiments, they can be considered as background electrons, which are not eliminated by those methods of background suppression, which were applied in experimental works.

In computer experiments the geometry of samples was the same, as in [11] (the thickness of a foil varied only), since in this work the results of measurement of electronic spectra in a wide energy range 0,1 MeV - 8 MeV are presented, that allowed to carry out the analysis of experimental and calculated spectra both for a high-energy and low energy part of spectra. As a result of computer experiments the electron spectra for samples of $^{235}$U, $^{239}$Pu and dioxide of $^{235}$U (93 % enrichment) representing a set of foils with a diameter 22 mm and thickness: 0,001 mm, 0,005 mm, 0,02 mm, 0,01мм, 0,1 mm, 1,0 mm. These foils were located between two protective aluminium disks of 0,025 mm thickness, in complete conformity with conditions experiment in [11], and the electron spectrum was fixed on the outcoming surface of a protective disk.

The comparative analysis of the experimental β-spectra [11], presented at fig. 2, we shall begin from a calculated spectrum received for last case: a foil $^{235}$U of dioxide of 93 % enrichment (diameter 22 mm and thickness 0,01мм), located between two protective aluminium disks of thickness 0,025 mm. The calculated spectrum is presented at the fig. 5.

In the table 4 and at the fig. 6 the relative errors which are brought in the measured electron spectrum by simulated processes of electron formation, not connected with reactions of β-decay and expressed in % are presented.

## 5. CONCLUSION

The analysis of errors (Fig. 6) shows that the processes of electron formation considered in the given work bring in the contribution in low energy part of experimentally measured β-spectrum. Since in a procedure of converting of the measured β-spectrum into the antineutrino spectrum the low energy part of β-spectrum is converted into the high energy part of antineutrino spectrum, the importance of such calculations and researches is seemed essential. Use of such simulating codes can allow to specify the β-spectra and as result the antineutrino spectra.

**REFERENCES**

[1] Rusov V.D., Vysotskii V.I., Zelentsova T.N. et al. The Neutrino Diagnostics of the Intareactor Processes and Fuel Containing Masses. // Nuclear and Rad. Safety. 1998., V.1. N.1, pp 66-95.

[2] Rusov V.D., Tarasov V.A., Tereshchenko D.F., Shaaban I. About one return task of neutrino diagnostics of intrareactor processes. // the Bulletin of the Kharkov university. Ser. Nuclei, fields, particles. 2002. №2. pp. 22-27.

[3] Rusov V.D., Zelentsova T.N., Tarasov V.O., Shaaban I. Statistical properties of the electronic antineutrino. // Reports of the NAS of Ukraine. 2002. №6. pp. 79-83.

[4] Beta- and antineutrino radiation of the radioactive nuclei: the directory/ Alexankin V.G., Rodichev S.V., Rubtsov P.M. et al.; Ed. Rubtsov P.M. - Moscow. Energoatomizdat, 1989. 800 p.

[5] Kopeikin V.I., Mikaelian L.A., Sinev V.V. The spectrum of electronic antineutrino of the nuclear reactor. // Nuclear physics. 1997. V. 60. pp. 230-234.

[6] Schreckenbach K., Colvin G., Gellety W., Feilitzsch F.V. Determination of antineutrino spectrum from $^{235}$U thermal neutron fission products up to 9.5 MeV. // Phys. Lett. 1985. Vol. 160B, №4. P. 325-330.

[7] Hahn A et al Antineutrino spectrum from $^{239}$Pu, $^{241}$Pu thermal neutron fission products // Phys. Lett. 1989. V.B218, P. 365-368.

[8] Vogel P., Schenter G.K., Mann F.M., Schenter R.E. Reactor antineutrino and their application to antineutrino-induced reactions II // Phys. Rev. 1981. V.C24. P.1543-1553

[9] Klapdor-Kleingrothaus G.V., Shtaudt A. Nonaccelerating physics of elementary particles. (In Russian). - M.: Science, Fizmatlit, 1997. 518 p.

[10] Mikaelyan L.A., Sinev V.V. Neutrino oscillations at reactors: what is next? // Nuclear Physics. 2000. T. 53, № 6. With. 1077-1081.

[11] Tsoulfanidis N. Wehring B. W., Wyman M. E. Measurements of time-dependent energy spectra of beta-rays from uranium-235 fission fragments. // Nucl. Sci. And Eng. 1971. Vol. 43. P. 42-53.

[12] Kolobashkin V.M., Rubtsov P.M., Ruzhanskii P.A., et al. The β-spectrum of fission products formed in nuclear reactor. // Nuclear physics. 1984. V. 40. №.2. p. 326-336.

[13] Carter R.E., Reines F., Wagner J.J., Wyman M.E. Free antineutrino absorption cross section. II. Expected cross section from measurements of fission fragment electron spectrum. // Phys. Rev. 1959. V. 113, №1. P. 280-286.

[14] Nakagawa T., Shibata K., Chiba S. et al. Japanese Evaluated Nuclear Data Library version 3 revision-2: JENDL-3.2 // J. Nucl. Sci. Technol. 1995. V.32. P.1259.

[15] Tasaka K., Takakura J., Ihara H. et al. JNDC Nuclear Data Library of Fission Products - Second Version, JAERI-1320, 1990.

[16] Tikhonov A.N., Arsenin V.Ya. Methods of the solution of incorrect tasks. M.: Science. 1979. 285 p.

Tikhonov A.N. et al. Regularizing algorithms and a priori information. M..:1983 183 C.

[17] Borovoi A.A., Kopeikin V.I., Mikaelian L.A., Tolokonnikov S.V. About the connection between the spectra of reactor $\tilde{v}_e$ and β-electrons // Nuclear physics. 1982. V. 36. pp 400-402.

[18] Kopeikin V.I., Mikaelyan L.A., Sinev V.V. Search for the neutrino magnetic moment in the nonequilibrium reactor- antineutrino energy spectrum // Phys. At. Nucl. 2000. V.63. P.1012.

[19] Radiating characteristics of the irradiated nuclear fuel: the directory / Kolobashkin V.M., Rubtsov P.M., Ruzhanskii P.A., SidorenkoV.D. M.: Energoatomizdat, 1983.

[20] Golubev B.P. Ionizing radiation dozimetry and protection // M.: Energoatomizdat. 1986. 461 p.

[21] Borovoi A.A., Klimov Yu.V., Kopeikin V.I. Measurement of electron spectra from $^{235}$U and $^{239}$Pu thermal neutron fission products. // Nuclear physics. 1983. V. 37. N.6. pp 1345-1350.

[22] Slavinskas D. D., Kennett T. J., Prestwish W. V. Resolution correction for β-ray spectra obtained with organic scintillators // Nucl. Instr. And Methods. 1965. Vol. 37. P. 36-41.

[23] Sen P., Patro A. P. Backscattering of electrons from phosphors and a matrix method for correcting backscattering and resolution in scintilation bete spectrometers // Nucl. Instr. And Methods. 1966. Vol. 40. P. 1-7.

[24] Kutcher J.M., Wyman M.E. An experimental study of the time dependence of the beta energy spectrum from 235U fission fragments. // Nucl. Sci. And Eng. 1966. Vol. 26. P. 435-446.

[25] Aleklett K., Nyman G., Rudstam G. Beta-decay properties of strongly neutron-rich nuclei. // Nucl. Phys. 1975. Vol. A246. P. 425-444.

[26] Rudstam G., Aleklett K. The energy distribution of antineutrinos originating from the decay of fission products in nuclear reactor. // Nucl. Sci. And Eng. 1979. Vol. 71. P. 301-308.

[27] Schreckenbach K. and e.a. Absolute measurement of beta spectrum from 235U fission as a basis for reactor antineutrino experiments. // Phys. Lett. 1981. Vol. 99B, №3. P. 251-256.

[28] Von Feilitzsch F., Hahn A.A., Schreckenbach K. Experimental beta-spectra from $^{239}$Pu and $^{235}$U thermal neutron fission products and their correlated antineutrino spectra. // Phys. Lett. 1982. Vol. 118B, № 1, 2, 3. P. 162-166.

[29] Kopeikin V.I. Electron and antineutrino spectra from products of fission of $^{235}$U, $^{239}$Pu, $^{241}$Pu by thermal and $^{238}$U by fast neutrons // Nuclear physics. 1980. V.32. pp 1507-1513.

Corresponding author:
V. Pavlovych
Prospekt Nauky, 47, Kyiv, 03680
Tel.: (380-44) 265-49-64; Fax: (380-44) 265-44-63
*E-mail: pavlovich@kinr.kiev.ua*

# FIGURES

Fig. 1. The ratio (p) of electron spectra of $^{252}Cf$ fission fragments for thick (imitation of uranium and plutonium targets) and thin sources.

Fig. 2. β-spectrum (electr/MeV·fis), measured in work [11] during 1000 s of thermal neutrons irradiation.

Fig. 3. The ratio $\rho_\beta^5(E)/\rho_\beta^9(E)$ of electron spectra from thermal neutron fission products of $^{235}U$ and $^{239}Pu$ [21]

Fig. 4. A beta-spectrum (el/MeV·fis), measured in works [28,29,6] for $^{235}U$

Fig. 5. Electron spectrum (el./MeV·fis), obtained by Monte-Carlo method for uranium dioxide of 93 % enrichment and protective aluminium disks.

Fig. 6. Relative errors (%) which are brought in the measured electron spectrum, obtained according to data of [11] and given work.

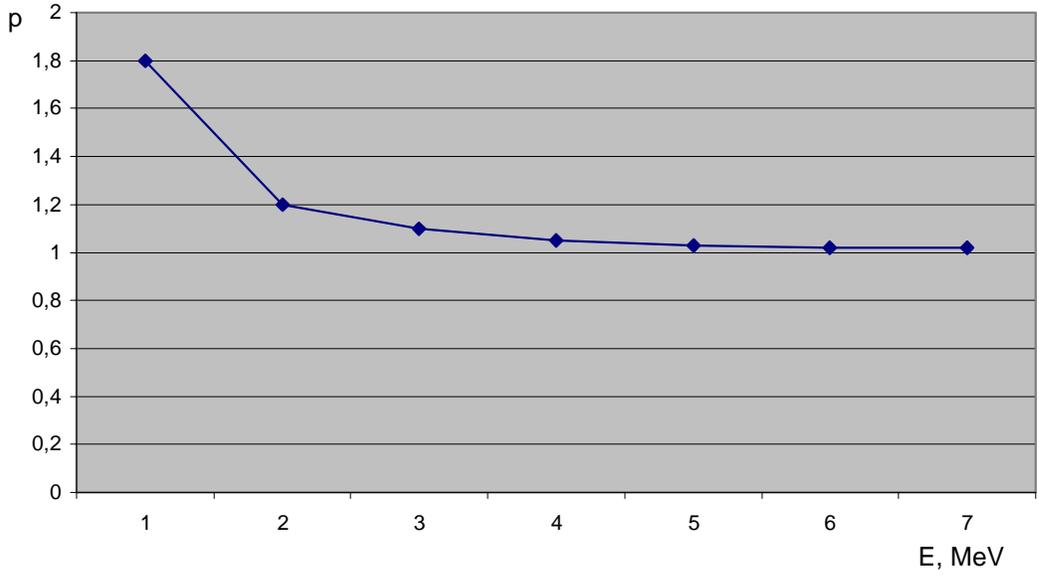

Fig. 1. The ratio (p) of electron spectra of $^{252}Cf$ fission fragments for thick (imitation of uranium and plutonium targets) and thin sources.

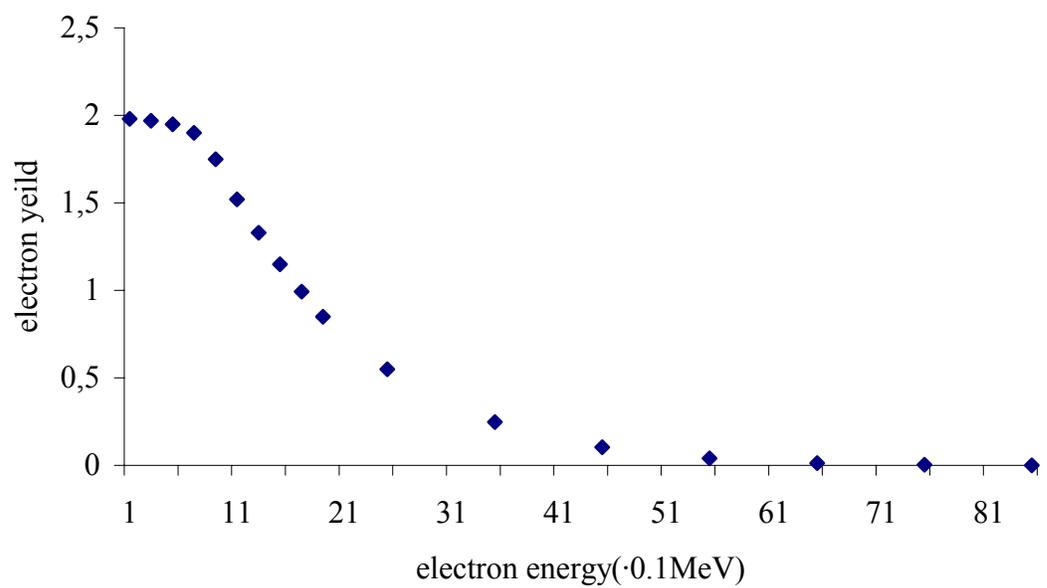

Fig. 2. β-spectrum (electr/MeV·fis), measured in work [11] during 1000 s of thermal neutrons irradiation.

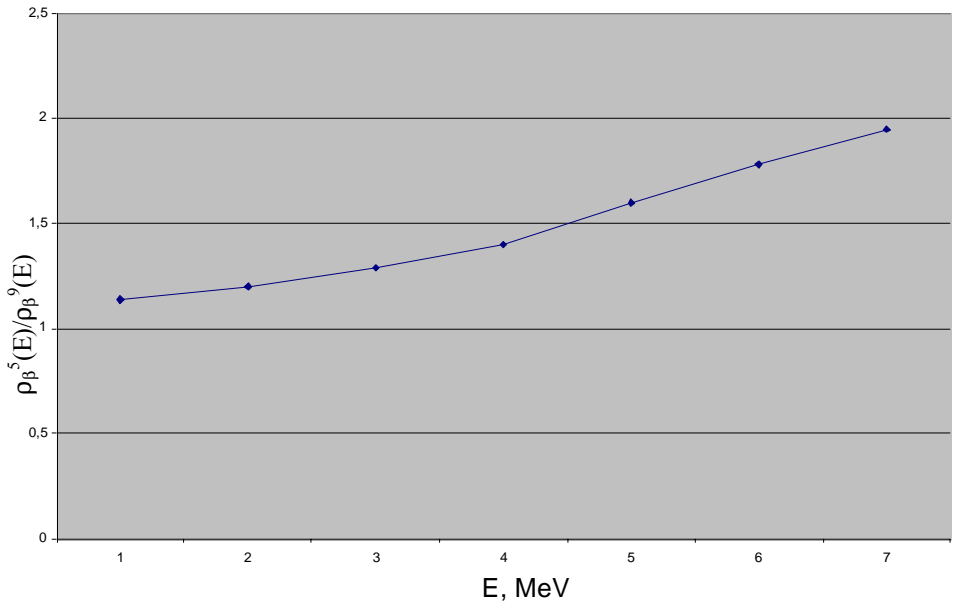

Fig. 3. The ratio $\rho_\beta^5(E)/\rho_\beta^9(E)$ of electron spectra from thermal neutron fission products of $^{235}U$ and $^{239}Pu$ [21]

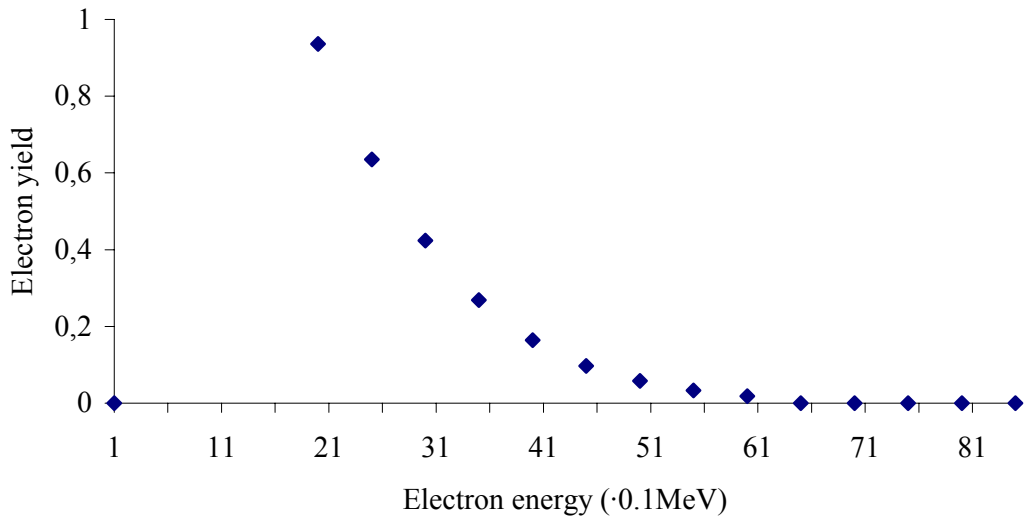

Fig. 4. A beta-spectrum (el/MeV·fis), measured in works [28,29,6] for $^{235}U$

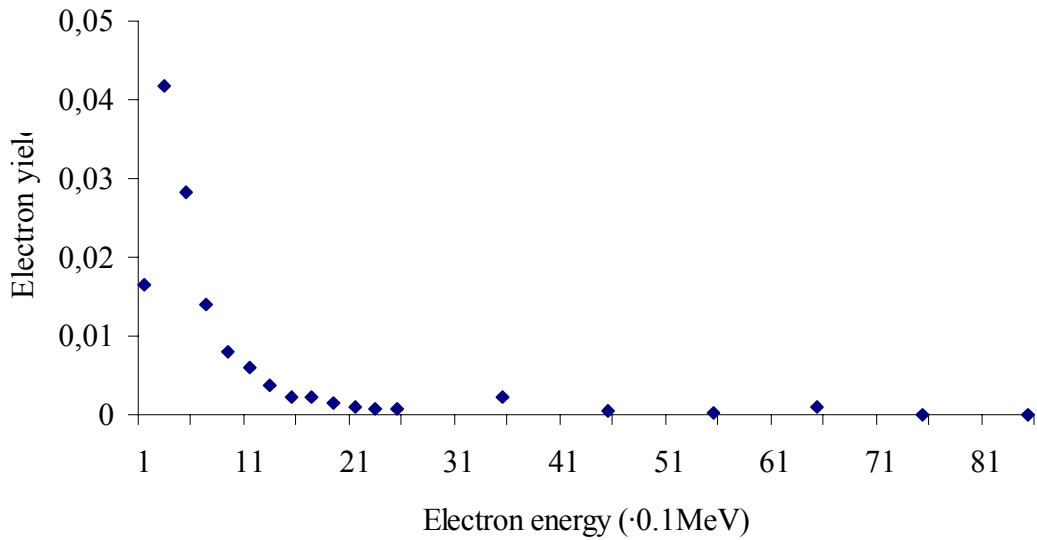

Fig. 5. Electron spectrum (el./MeV·fis), obtained by Monte-Carlo method for uranium dioxide of 93 % enrichment and protective aluminium disks.

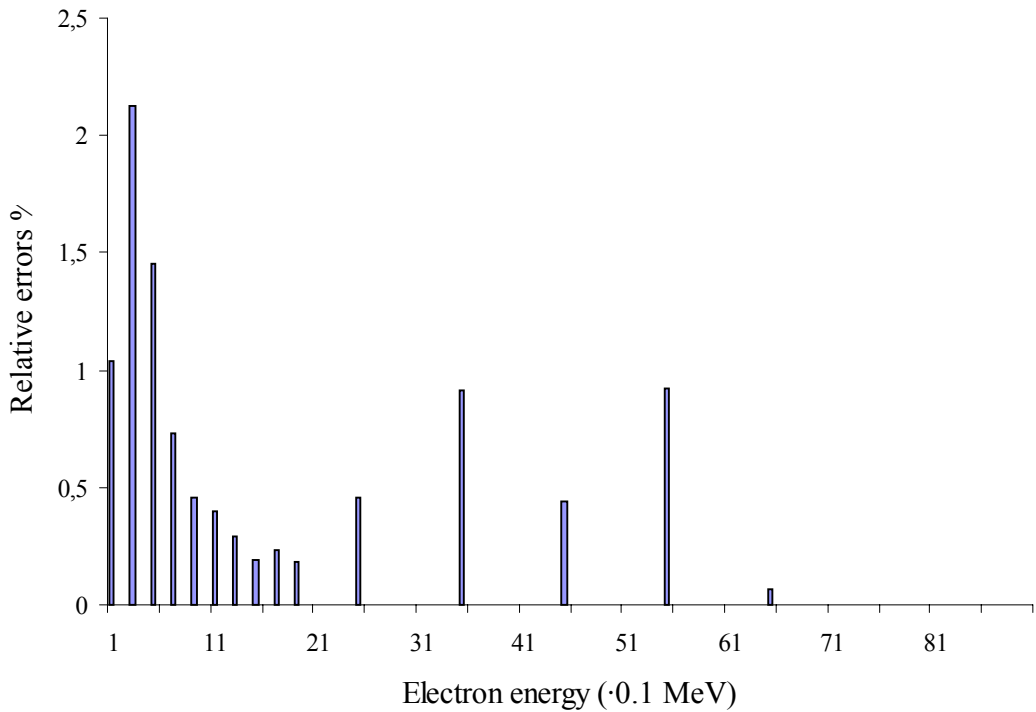

Fig. 6. Relative errors (%) which are brought in the measured electron spectrum, obtained according to data of [11] and given work.